\documentclass[twocolumn,aps,superscriptaddress,longbibliography]{revtex4-2}
\usepackage{amsmath,amssymb,graphicx}
\usepackage[utf8]{inputenc}
\usepackage[colorlinks=true, allcolors=blue]{hyperref}

\begin{document}

\title{Computational foundations of the human world}

\author{Marcus~J.~Hamilton}
\email{marcus.hamilton@utsa.edu}
\affiliation{Department of Anthropology, University of Texas at San Antonio, San Antonio, TX 78249, USA}
\affiliation{Center for Data Science, University of Texas at San Antonio, San Antonio, TX 78207, USA}
\affiliation{Santa Fe Institute, Santa Fe, NM 87501, USA}

\author{Abhishek~Yadav}
\affiliation{Santa Fe Institute, Santa Fe, NM 87501, USA}
\affiliation{Department of Computer Science, University of New Mexico, Albuquerque, NM 87131, USA}

\author{Harrison~Hartle}
\affiliation{Santa Fe Institute, Santa Fe, NM 87501, USA}

\author{Jan~Korbel}
\affiliation{Complexity Science Hub, Vienna, Austria}

\author{Niels~Kornerup}
\affiliation{Sandia National Laboratories, Albuquerque, NM 87123, USA}

\author{Andrew~J.~Stier}
\affiliation{Santa Fe Institute, Santa Fe, NM 87501, USA}

\author{Douglas~H.~Erwin}
\affiliation{Santa Fe Institute, Santa Fe, NM 87501, USA}

\author{Hyejin~Youn}
\affiliation{Santa Fe Institute, Santa Fe, NM 87501, USA}
\affiliation{Seoul National University, Korea}

\author{Christopher~P.~Kempes}
\affiliation{Santa Fe Institute, Santa Fe, NM 87501, USA}

\author{Hajime~Shimao}
\affiliation{Great Valley School of Professional Studies, Penn State, Malvern, PA 19355, USA}

\author{Kyle~Harper}
\affiliation{Santa Fe Institute, Santa Fe, NM 87501, USA}

\affiliation{Department of Classics and Letters, University of Oklahoma, Norman, OK 73019}

\author{James~Evans}
\affiliation{Santa Fe Institute, Santa Fe, NM 87501, USA}

\affiliation{Department of Sociology, University of Chicago, Chicago, IL}

\author{David~H.~Wolpert}
\affiliation{Santa Fe Institute, Santa Fe, NM 87501, USA}
\affiliation{Complexity Science Hub, Vienna, Austria}
\affiliation{Arizona State University, Tempe, AZ 85281, USA}
\affiliation{International Center for Theoretical Physics, Trieste 34151, Italy}

\begin{abstract}
Human societies continuously transform scattered information into collective judgments and coordinated action, whether through markets discovering prices, governments allocating resources, communities enforcing norms, or science converging on reliable claims. Importantly, the computational difficulty of collective decision-making, particularly the time and communication required to reach solutions, imposes fundamental constraints on social organization. While theoretical computer science offers formal tools for analyzing such problems, for instance, by analyzing resource requirements, including time and memory, surprisingly, there is no domain of social science that focuses on the nature of computation in the human world. This perspective argues that we now have the opportunity to deploy these computational frameworks to study human social organization, opening research directions at the intersection of computer science and social science. We highlight core social phenomena that can be framed as computational, including (i) distributed consensus and coordinated action, (ii) societal restructuring with scale, (iii) hierarchical and modular structure, and (iv) externalized memory systems. We identify several concepts from theoretical computer science that may provide insight into these phenomena, especially emphasizing more recently developed approaches beyond the paradigm of Turing~Machines and worst-case computational complexity. 
\end{abstract}

\maketitle

\section{Introduction}

For physical, technological, biological, and social systems, information processing is fundamental. A longstanding aspiration in the natural and social sciences has been to understand how large populations of interacting individuals might admit effective macroscopic descriptions~\cite{Ball2004,castellano2009statistical}.
The quantitative study of human societies has led to empirical and theoretical advances in social network science \cite{Granovetter1973,Onnela2007,Centola2010}, econophysics \cite{Gabaix2003,Plerou2003}, human-centered experimental game theory \cite{Fehr2000,Rand2011}, scaling and allometric approaches \cite{West1997,barabasi2005origin,Bettencourt2007,Bettencourt2013}, and quantitative historical dynamics \cite{Turchin2018,Shin2020,Turchin2022}. While substantial progress has been made in identifying statistical regularities \cite{gell2011regularities}, scaling relations \cite{rybski2009scaling}, and recurrent patterns, a unified framework connecting microscopic social interactions to robust macroscopic behavior has so far proven elusive; on the whole, the application of formal models and theories to the study of human social systems remains fragmented and \textit{ad hoc}. In part, this reflects the nature of the subject: there is very little data for much of human history, and what data there is tends to be noisy, sparse, and biased. However, we suggest this gives all the more reason for new theoretical approaches with a strong footing in more mathematically rigorous concepts.

\begin{figure}[!h]
\centering
\includegraphics[trim=10 5 0 0,width=1.0\linewidth,clip]{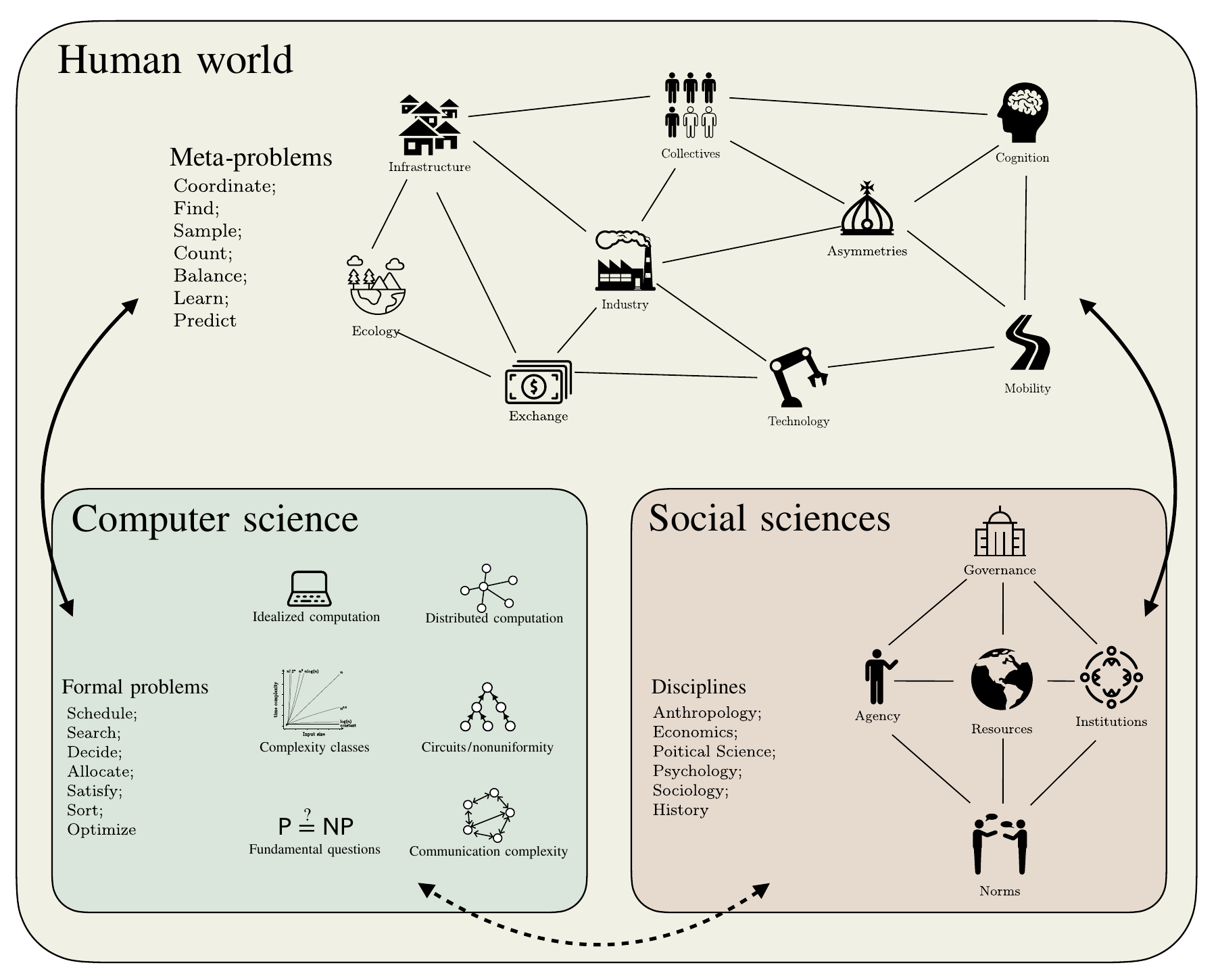}
\caption{The human world confronts a variety of general ``meta-problems'' such as finding, counting, coordination, prediction, and so on. Computer science formalizes problems such as scheduling, searching, decision making, optimization, etc. Most abstractly, these questions are addressed in the setting of Turing machines and worst-case complexity, which formalize the fundamental nature of computation. Less idealized notions such as non-uniform complexity and distributed computing formalize constraints and structure of problem solving in real world architectures, such as Boolean circuits, capturing modularity and hierarchy in computation. Meanwhile, the social sciences observe the human world and its meta-problems and are more often concerned with the empirical nature of problem solving in specific domains---traditionally focusing on traits such as norms, institutions, agency, and governance. Here, we simply propose to close the conceptual loop by  linking social sciences to computer science theory.}
\label{fig:schematic}
\end{figure}

In this Perspective, we propose that since all human social systems process a vast amount of information at multiple levels to determine how to coordinate and organize action~\cite{wolpert2024computational}, the mathematics of information processing could offer a unifying framework to understand social dynamics. In particular, we propose here that computer science theory might be a fruitful standpoint for the investigation of the principles underlying social systems, especially when integrated with traditional and quantitative social science approaches. Identifying ``meta-problems'' that human social collectives seek to solve, and mapping the variables of those settings to problems that are rigorously examined in computer science, one arrives at a computational lens through which to examine social processes (Figure~\ref{fig:schematic}).

\begin{figure}[!h]
\centering
\includegraphics[width=1\linewidth]{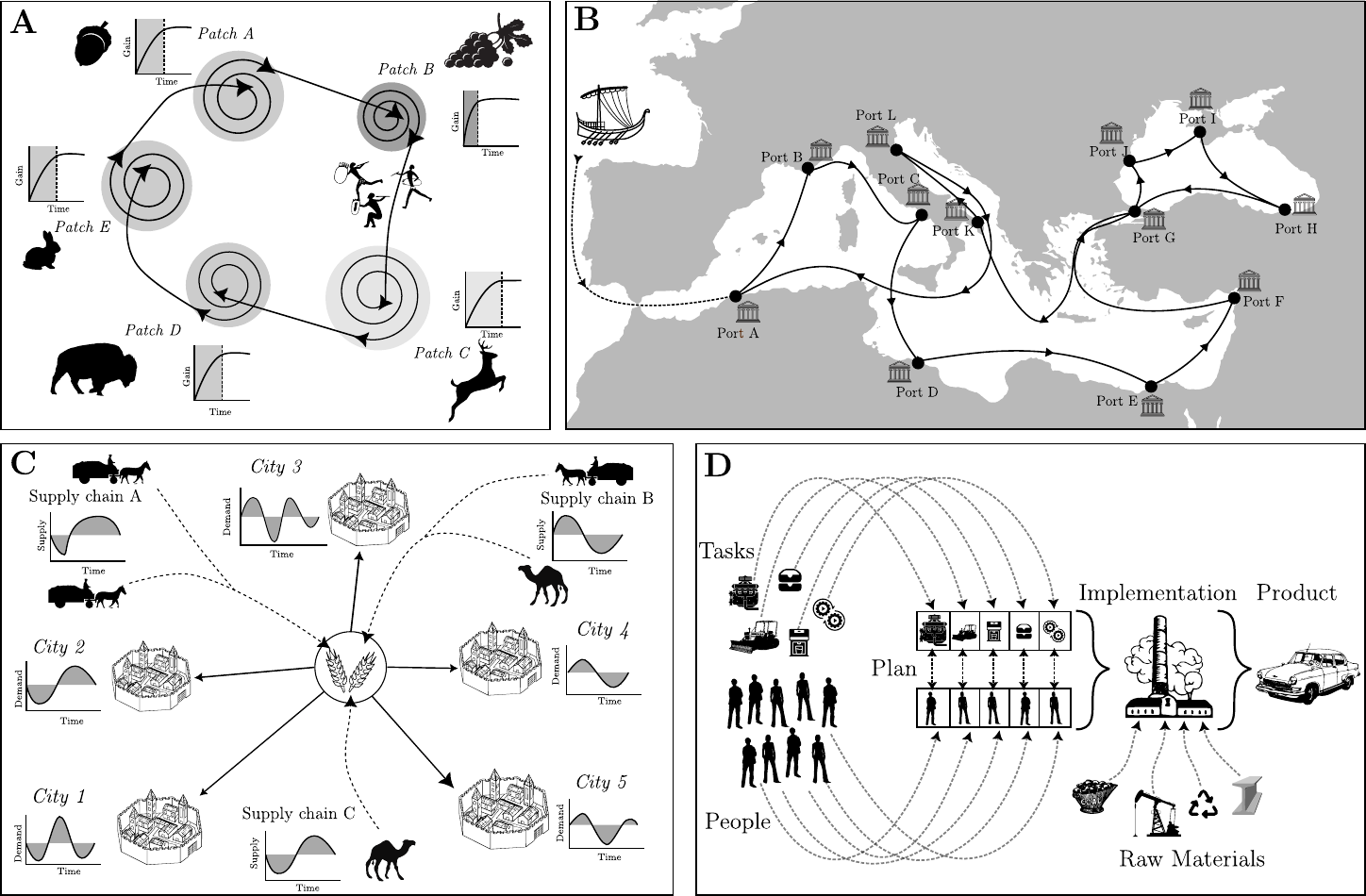}
\caption{Computational abstractions of meta-problems in the human world. A) Time optimization problems, such as patch residence time in foraging theory, where the decision is how long to stay in a patch given known alternatives; B) mobility/traveling salesperson-like problems, where multiple locations must be visited under constraint; C) allocation problems, where the information of individual supply and demand needs to be processed for redistribution of resource among various entities; and D) scheduling problems, where multiple inputs must be coordinated to manufacture an end product.}
\label{fig:4_problems}
\end{figure}

Broadly speaking, computational systems map input information to decisions or outputs through defined procedures, and many naturally occurring systems can be viewed as computing~\cite{wolpert2024stochastic,wolpert2026does}. In this sense, human societies are distributed computational systems, aggregating dispersed information, knowledge, and skills to solve problems of collective behavior \cite{brush2018conflicts,hamilton2022collective,wolpert2024computational}. These problems broadly include resource allocation, collective decision-making, and integration of specialized skills and occupations to efficiently create goods or services~\cite{arrow1964social, durkheim2019division}. These core problems (Figure~\ref{fig:4_problems}) are computational in nature, and arise unavoidably when groups form. Such problems are extremely general and have deep evolutionary histories, arising in small-scale societies and modern nation-states alike. Social choice theory formalizes many of these challenges as mathematical aggregation functions~\cite{sen1977social}.

Coordinating competition for scarce resources and allocating them among possible uses are fundamental problems faced by social systems~\cite{beinhocker}. Hayek famously argued that the market functions as an information-processing system of some ill-specified sort~\cite{hayek1945use}, or, in modern terms, as a computational process. Along the same lines, institutions and bureaucracies function to coordinate specialized skills and knowledge to produce complex goods and services~\cite{north1989institutions, north1990institutions}. For the vast majority of human evolutionary history, human societies self-organized into small-scale, quasi-egalitarian communities, managing local adaptive problems in groups of perhaps no more than a few dozen individuals~\cite{dunbar2007evolution, hamilton2022collective, hamilton2007complex}. As societies grew, they developed increasingly specialized roles and institutional infrastructures to coordinate internal dynamics~\cite{bowles2009microeconomics, hamilton2024institutional}; arguably, these restructurings provide greater {\it computational} power, enabling societies to solve coordination problems at previously unimaginable scales~\cite{kohler2022social, Shin2020}. These trends suggest that the way societies process information for decision making, and how efficiently they do so, constrains their growth and dynamics.

Theoretical computer science has developed a rich set of tools for the systematic and rigorous analysis of problem hardness and the efficiency with which procedures solve them~\cite{moore2011nature, arora2009computational, lynch1996distributed, papadimitriou2003computational, papadimitriou2013combinatorial}. These tools provide a powerful lens for understanding the computational aspects of collective behavior in social systems. In particular, subfields such as distributed computation and approximation complexity offer methods with direct and realistic applicability. In this Perspective, we outline a research agenda that brings together existing work at the boundary of computation and social systems. 
% To disambiguate our viewpoint, first we outline how our approach differs from related yet distinct research directions in Sec.~\ref{ssec:related_scope}. 
We begin by presenting 
% We then proceed with 
our main points, arguing that many societal functions are fundamentally computational (Sec.~\ref{sec:attributes}). Then we describe how current tools from computer science can provide insight into this computational nature (Sec.~\ref{sec:cs})---particularly when extended beyond traditional models and measures of computation, such as the Turing~Machine and worst-case complexity analysis. We survey some related research in Sec.~\ref{sec:related}. In particular, we clarify the scope of our Perspective in Sec.~\ref{ssec:related_scope}, distinguishing 
it from (i) the use of computation in studying social phenomena (computational social science), (ii) the study of algorithms {\it in} society, and (iii) the design of algorithms for societal benefit; our focus instead is on the computational aspects of social phenomena. We conclude in Sec.~\ref{sec:conclusion} by identifying key challenges and sketching a path toward developing a computational foundation for the study of the human world.

\section{Computational attributes of human social systems}
\label{sec:attributes}

From hunter-gatherer populations to industrialized economies and scientific enterprises, societies engage in large-scale information processing: they solve problems, make decisions, and encode, accumulate, and deploy knowledge across generations~\cite{hamilton2022collective, boyd1988culture, boyd2011cultural, Hamilton2025MoreIsDifferent, mokyr2016culture}. In this section, we identify key computational aspects of social coordination.

\subsection{Centralized versus distributed information and computation}
\label{ssec:coordination}

To delineate the distributed and/or centralized nature of computation in human collectives, it is important to note a crucial distinction concerning where information required for decision-making resides versus where computations are performed. The information required for collective decision-making is almost always distributed throughout a population, whereas computation can range from fully centralized to fully distributed~\cite{hayek1945use}.

Cases of centralized computation are common in social systems. For example, in democratic elections, information about individual preferences is distributed across the electorate, collected and aggregated by a central, trusted election commission, and then used to determine an outcome. Both the method of collecting preferences and the rule for computing the result vary widely, from plurality and ranked-choice voting to approval voting and other systems~\cite{jackson2014mechanism}, but the computation itself is centralized. Similarly, in auctions where information about private valuations is dispersed across bidders, a central auctioneer collects this information to determine the winner via a predetermined procedure
e.g., sealed-bid second-price auctions or ascending-bid formats. Many other problems can also be viewed similarly, such as public policy selection~\cite{jackson2001crash}.

Fully decentralized systems represent the opposite extreme. For instance, in an idealized free-market economy, supply and demand information is dispersed heterogeneously across buyers and sellers, and prices emerge through decentralized trading. More broadly, many self-organizing aspects of economies can be categorized as distributed computation~\cite{van2017distributed, krugman1996self}. In the words of Hayek, ``{\it the information required to solve these problems of the whole society is never given to a single mind which could work out the implications and can never be so given. The economic problem of society is a problem of the utilization of knowledge which is not given to anyone in its totality}''~\cite{hayek1945use}. Other examples include the emergence of languages, cultural norms, and narratives. Recent examples may include cryptocurrency, decentralized governance schemes, and online social networks.

Most organizational forms, such as military decision making, management of cities and states, and even bureaucratic governance, lie between these extremes. For instance, military structures illustrate how information gathered at the tactical level is passed upward through successive layers of command, and the strategic decisions made at higher levels flow back down the hierarchy as executable orders. Bureaucratic governance, corporate management, and large-scale logistics networks similarly employ hierarchical computation---partially distributed (local autonomy at lower levels) yet coordinated through higher-level integration. 

There has been much work on the ``efficiency'' of outcomes arrived at by these different systems, whether resources are allocated optimally (Pareto efficiency), or whether collective decisions maximize social welfare~\cite{jackson2001crash, jackson2014mechanism}. But this raises a fundamental question; can outcomes that are socially optimal be computed efficiently in the first place? If computing a socially optimal allocation requires exponential time, astronomical communication costs, or unwieldy computational power, then the efficiency of that theoretical optimum becomes irrelevant to real-world situations. For instance, a long tradition in economics has argued that decentralized markets achieve economically efficient allocations that would be computationally infeasible for a centralized mechanism~\cite{von2024economic, arrow1974limits, hayek1945use}. More generally, the computational feasibility of reaching outcomes, not just their desirability, is crucial for the study of social systems.

\subsection{Hierarchical and modular computation}
\label{ssec:mod_hier}

Real-world organizational forms typically lie somewhere between the theoretical extremes of fully centralized and fully distributed. Patterns of human social connectivity have been studied widely \cite{ClydeMitchell,marin2011social}, with salient features including high clustering coefficient \cite{newman2003social} (often modeled as emerging through triadic closure \cite{TOIVONEN2006851,holme2002growing}), small-worldness \cite{watts1999small,watts1998collective}, and highly heterogeneous numbers of contacts \cite{barabasi1999emergence}. In this section, we emphasize two additional properties of human social collectives: modular and hierarchical structure.

\subsubsection*{Modularity} In essence, a modular system is composed of multiple semi-independent components. In human collective behavior, modularity allows different components---e.g., individuals, groups, or institutions---to specialize in distinct domains and handle different problem aspects. In governance, federal systems create modular units: states manage education and local infrastructure while the national government handles defense and foreign policy. In production, division of labor creates specialized roles: one worker masters circuit assembly, another manages supply chains, neither needing the other's detailed expertise. In organizations, departments operate semi-autonomously: the marketing team doesn't need to understand the manufacturing processes in detail. In social networks, modularity indicates {\it community structure} \cite{girvan2002community}, groupings of nodes for which connectivity within groups is dense and between groups is sparse \cite{newman2004finding,lancichinetti2009community}. The sociology of community has been subject to much investigation \cite{wellman1979community}.

\subsubsection*{Hierarchy} Qualitatively, hierarchy refers to a branching, layered interdependency~\cite{simon2012architecture}. Functionally, hierarchical roles allow the output of different modules to be integrated across different levels; lower levels possess detailed local knowledge and make routine decisions autonomously, whereas higher levels integrate information across modules and address broader coordination challenges. This fractal-like structure appears throughout human societies: from local governance to national policymaking, from corporate departments to military command chains, and across traditional non-industrialized non-egalitarian societies~\cite{hamilton2020scaling}. A variety of measures of hierarchy have been defined \cite{trusina2004hierarchy,eum2008new,mones2012hierarchy,Krackhardt1994,EVERETT2012159}; hierarchical organization has been examined in human and nonhuman social systems \cite{shizuka2012social,gupte2011finding, 10.1098/rstb.2020.0440, shively1985evolution,sapolsky2005influence,boehm2009hierarchy}.

Hierarchical and modular organization is often observed in human social systems, especially in larger populations~\cite{perret2020disorganized}. For millennia, hunter-gatherer groups of a few dozen people operated with minimal hierarchy~\cite{hamilton2007complex}. Small-scale agricultural communities of perhaps hundreds of people often developed societies with two-tier sociopolitical hierarchies \cite{service1962primitive,earle1997chiefs}. City-states, composed of thousands of people, required many more bureaucratic layers \cite{Schott01012000}. Contemporary nation-states made up of millions of people often employ deep hierarchical bureaucracies with extensive modularization---for instance, federal agencies, state governments, and municipal departments, each with internal hierarchies~\cite{hamilton2020scaling}. Similarly, production evolved from individual craftsmen to factories and modern supply chains via increasing specialization of roles and division of labor, as coordinated through multiple hierarchical layers~\cite{maccarthy2016supply}.

Together, modularity and hierarchy have been argued to make the problem of utilizing and aggregating existing knowledge far more manageable~\cite{simon2012architecture}. But why are they commonly observed in social systems? This has been a central question in sociology  \cite{chase1980,magee2008} and in the study of social networks \cite{yang2008,marcoux2013network,lu2016}. We suggest that a computational perspective could provide complementary insights by framing human organizations as solving problems by integrating dispersed knowledge to coordinate actions. We may then evaluate social structure in reference to the associated abstract computational problems. This motivates the development of formal models addressing how organizational structure scales with population size and knowledge complexity; what communication costs different architectures impose; when can hierarchies improve or hinder coordination; how hierarchical and modular structures affect robustness to failures.

\subsection{Scaling of human social computation}
\label{ssec:scaling}

As human populations grow in scale, they generate more information to process and knowledge to utilize, but also face coordination problems that are fundamentally harder to solve. A group of a few dozen hunter-gatherers coordinating a hunt face qualitatively different challenges to a city of many thousands coordinating food distribution, or a nation of millions coordinating economic production. Mounting coordination costs impose significant challenges involving growing conflicts of interest, potentially leading to computational intractability and combinatorial blowup.

Yet, interestingly, rather than succumbing to these limits, human societies repeatedly reorganize \textit{how} they compute. Larger populations may be expected to provide increased information processing capacity simply by having more people. But, as populations grow, they reconfigure, altering the structure of institutions, communication pathways, forming divisions of labor, and creating new collective functions capable of confronting more complex challenges \cite{jeanson2007emergence}. As such, the relationship between problem size and problem-solving capacity co-evolves over human history, contributing to the diversity of cultural, institutional and organizational forms we observe today.

Moreover, research on scaling in human societies reveals consistent patterns of growth and self-organization, showing as societies expand, they systematically transform to become more efficient, according to several metrics~\cite{west2018scale,bettencourt2021introduction,hamilton2020scaling}. For instance, cities display superlinear scaling in productivity as larger populations yield disproportionately more innovation, economic output, and knowledge creation \cite{Arbesman2009,Bettencourt2013}. We suggest that at least in part, these phenomenological scaling behaviors reflect the increased computational demands of information processing and collective coordination. It would be fruitful to develop formal frameworks addressing the communication costs of different organizational architectures across scales. How does problem difficulty scale with increasing population size? When and how do such restructurings provide computational benefits?

\subsection{Knowledge accumulation, encoding, and external information processing}
\label{ssec:memory}

One of the striking capacities of human societies is the ability to extend computational capacity through external memory systems \cite{clark1998extended, clark2008supersizing, donald1993origins}. This allows societies to accumulate, encode, and deploy knowledge across generations, reducing the need to recompute solutions to problems that have already been solved \cite{henrich2003evolution, Hamilton2025MoreIsDifferent, mokyr2016culture}. 

For example, as human societies developed increasingly complex institutions, organizations, and technology, they streamlined writing systems~\cite{schmandt2014evolution}. Writing transformed human adaptive computation by enabling the external storage of knowledge, increasing societies’ capacity to process information \cite{martin1994history}. As such, external memory systems such as writing function not only as facilitators of exchange~\cite{mcwilliams2024money}, but also as repositories of past decisions, compressing adaptive knowledge into robust forms and stabilizing its transmission across generations.

The recent transition to digital knowledge storage~\cite{Cheng2023} and AI-assisted decision-making~\cite{Steyvers2024} is merely the latest expansion of human adaptive computational ability, allowing real-time access to vast amounts of information and enhancing predictive capabilities. But such transitions have a deep evolutionary history \cite{Foley2016}.

\section{Computer science theory relevant for human social systems}
\label{sec:cs}

Computational complexity theory emerged to formally characterize the intrinsic difficulty of problems, providing a framework for comparing the efficiency of algorithms via {\it complexity classes}, defined based on the resources---such as time, memory, or communication---required by the algorithm as a function of input size \cite{Knuth}. A problem is considered {\it efficiently computable} if solvable by a conventional Turing~Machine ``quickly'', that is, if it resides in class $\mathsf{P}$: the class of decision problems solvable in polynomial time. Some examples of problems known to be in $\mathsf{P}$ include sorting an array,  multiplying matrices, and finding shortest paths in a graph~\cite{knuth1997art}. This efficiently computable class is conjectured to be distinct from class $\mathsf{NP}$, problems whose solutions can be \textit{verified} in polynomial time but not necessarily found as quickly. The hardest problems in $\mathsf{NP}$ are called $\mathsf{NP}$-complete; if any $\mathsf{NP}$-complete problem has a polynomial-time algorithm, then so too must all other problems in $\mathsf{NP}$. Since no polynomial time algorithms are known for $\mathsf{NP}$-complete problems, they are often considered intractable. Other examples of $\mathsf{NP}$-complete problems include the traveling salesman, knapsack, and longest path problems~\cite{karp2009reducibility}.  Perhaps the most quintessential $\mathsf{NP}$-complete problem is Boolean satisfiability, which was the first problem proven to be $\mathsf{NP}$-complete \cite{Cook71}: given a sequence of clauses, does there exist an assignment of Boolean variables that satisfy them all simultaneously? Whether $\mathsf{P}=\mathsf{NP}$, that is, whether efficient verification implies efficient solution-finding, remains the most fundamental open question in theoretical computer science~\cite{Cook71,aaronson201310}.

\subsection{Turing Machine}
\label{ssec:TM}

Central to the definitions of $\mathsf{P}$ and $\mathsf{NP}$ is the Turing~Machine (TM), an idealized model of computation that consists of an infinitely long tape that can be read from and written to, a state register, and an instruction table, allowing for execution of a deterministic, sequential algorithm~\cite{arora2009computational}. The Church-Turing thesis~\cite{church1936unsolvable,turing1936computable} holds that a Turing~Machine is a universal model of computation, meaning that it is theoretically capable of simulating any algorithm or computing device, thereby defining the absolute limits of what is computable. 

However, a Turing~Machine does not reflect the distributed, stochastic, and often parallel nature of computation in human collectives. Therefore, it does not necessarily provide an appropriate framework for analyzing the unique challenges that arise in the noisy, stochastic, distributed computational environments of the real world.

\subsection{Non-uniform computation}
\label{ssec:circuits}

In uniform models of computation, like the Turing machine, a single computational procedure applies to inputs of all sizes. On the other hand, {\it non-uniform} models of computation allow the computational procedure to vary with the input length. This distinction parallels the structural reconfiguration undergone by human collectives as their scale changes, as discussed in Sec.~\ref{ssec:scaling}. Below, we describe two standard formulations of non-uniform computation.

\subsubsection*{Boolean circuits} Boolean circuits are directed acyclic graphs \cite{thulasiraman2011graphs} where each node represents a logic gate (AND, OR, NOT, etc.). A circuit computes a Boolean function $f:\{0, 1\}^n \to \{0, 1\}^m$ by mapping $n$ input bits through the intermediate gates to produce $m$ output bits. Boolean circuits characterize computational difficulty through two key measures: size (number of gates) and depth (length of the longest path from input to output). Circuits naturally model {\it parallel} computation, where gates within the same layer can execute simultaneously. Nick's Class ($\mathsf{NC}$) is defined as functions computable by polynomial-size, polylogarithmic-depth circuits, characterizing problems efficiently solvable in parallel. More precisely, $\mathsf{NC}^i$ consists of functions computable by polynomial-size circuits with constant fan-in and depth $O(\log^i n)$, and $\mathsf{NC} := \bigcup_{i=0}^{\infty} \mathsf{NC}^i$. For example, computing PARITY or MAJORITY of $n$ bits belong to $\mathsf{NC}^1$~\cite{arora2009computational}.

Circuits can explicitly embody modular and hierarchical computation (Sec.~\ref{ssec:mod_hier}). Each gate represents a distinct, localized computation, the output of which can be shared as input for multiple subsequent gates. This structure intrinsically highlights the reuse of computational results and the building of complex operations from simpler, discrete modules, and more closely resembles the way computation in human groups appears to be organized.

\subsubsection*{Advice} Non-uniform computation allows for different computational architectures at different input lengths, but does not account for the cost of determining the appropriate procedure at each input size. This abstraction is formalized through the notion of advice~\cite{arora2009computational}, whereby size-dependent but instance-independent information supplements the computation. Complexity class $\mathsf{DTIME}(T(n))/a(n)$ denotes problems decidable by a TM in $T(n)$ steps given access to optimally chosen advice strings of length $a(n)$. In the context of Boolean circuits, such advice can be understood as specifying the optimal circuit architecture for inputs of length $n$.

Conceptually, the notion of advice resembles how collective wisdom, culture, knowledge, and norms can allow problems to be solved more easily than in the absence of such information. New organizations do not form entirely from scratch; they instead rely on large swaths of information, often encoded in external memory systems as discussed in Sec.~\ref{ssec:memory}. However, the analogy between useful cultural information and advice strings is limited in that the latter are typically defined as theoretically optimal. Instead, more relaxed notions such as randomized~\cite{komm2011advice,bockenhauer2017online} or untrustworthy~\cite{angelopoulos2024online} advice strings are more analogous to the human setting. The notion of advice also mirrors the behavior discussed in Sec.~\ref{ssec:scaling}, whereby human societies restructure with scale.

\subsection{Complexity paradigms}
\label{ssec:complexity_notions}

A predominant paradigm for comparing the efficiency of algorithms is based on worst-case computational complexity: the scaling of runtime required to solve problems given the most adversarial choice among all possible inputs. Worst-case analysis plays a foundational role in formal models of computation because it offers a mathematically tractable way to reason about computational difficulty. This framing underlies the classical definitions of complexity classes such as $\mathsf{P}$ and $\mathsf{NP}$. Worst-case analysis provides an analytical way to compare the efficiency of two different algorithms: a ``better'' algorithm is one with better worst-case performance. Worst-case computational complexity captures a specific notion of problem difficulty, but other notions of complexity are also well-motivated, as we discuss below.

\subsubsection*{Average-case complexity} 

In some practical scenarios, worst-case analysis is shown to be misleading: even though algorithms have polynomial runtime for all practical purposes, worst-case analysis predicts exponential runtime~\cite{roughgarden2019beyond}. For instance, almost all pairs of graphs can be tested quickly for isomorphism \cite{babai1980random}, despite graph isomorphism having no known polynomial-time algorithm for worst-case instances. This motivates notions of {\it average-case} complexity where a probability distribution over inputs is provided \cite{levin1986average}. Average-case complexity allows characterization of problem difficulty for ``typical'' instances, rather than adversarially chosen worst-case inputs~\cite{bogdanov2006average}. More closely related to social phenomena, Ben-Or's probabilistic consensus algorithm \cite{ben1983another} is designed for scenarios with probability distributions over initial states, as opposed to a worst-case formulation. 

\subsubsection*{Approximation complexity} 
\label{sssec:approx}

Approximation complexity is a framework that focuses on the difficulty of finding solutions that are ``good enough'' relative to the optimum, rather than requiring exact solutions in the worst case~\cite{papadimitriou1988optimization}. This paradigm captures computational strategies that balance resource costs, uncertainty, and the need for reliable but not necessarily perfect outcomes. The standard formulation of approximation complexity \cite{vazirani2001approximation} asks: how much time does it take to approximate an optimal solution within a constant factor? Some problems have been shown to be {\it inapproximable} in that even finding approximate solutions is NP-hard \cite{hochba1997approximation}.

While classical complexity theory often emphasizes the required resources for an exact computation, real social systems often operate under severe time and resource constraints that make exact optimization infeasible \cite{Haenni2002}. This motivates quantifying the best quality of approximation achievable given finite time, memory, or energy budgets~\cite{vazirani2001approximation}. The corresponding computational questions include: what approximation guarantees are achievable under realistic constraints? How does organizational structure---e.g., in terms of hierarchical versus distributed---affect the speed and quality of convergence toward approximate solutions?

\subsection{Distributed computing theory}
\label{ssec:distributed}

A more recently developed branch of computer science is distributed computing theory~\cite{van2017distributed}, which studies how multiple computing units can coordinate to solve problems under various constraints. A distributed system consists of a graph with nodes representing processors and edges representing communication channels. A distributed algorithm assigns each node a program, enabling it to perform local computations and exchange messages with neighbors. In distributed computing tasks, in addition to the time required for computation, the amount of communication required becomes critical~\cite{kushilevitz1997communication,peleg2000distributed}. A high communication cost can make a distributed procedure impractical to implement and scale. Another challenge is the effect of asynchrony: if events are not governed by a single clock and don't occur at precisely known times, collective computations can become more difficult~\cite{lamport2019time}. Models of distributed computation involve design choices for communication and timing mechanisms~\cite{lynch1996distributed}. This flexibility allows distributed computing theory to formalize and study diverse aspects of collective computation in groups of agents.

Many results on the performance of distributed computing systems have been obtained. For instance, it has been shown that distributed algorithms cannot simultaneously guarantee three desiderata: consistency, availability, and partition tolerance---capturing fundamental trade-offs in system design \cite{brewer2000towards,lynch1996distributed}. Communication complexity~\cite{yao1979some} characterizes the minimum information exchange required for distributed parties to compute functions. For instance, if two parties each have an $n$-bit number, at least $n$ bits of a deterministic communication protocol are required to determine whether they are equal~\cite{Roughgarden2016}. Associated communication complexity classes have been defined \cite{babai1986complexity}. Two-party settings are the most common~\cite{nisan1993communication}, but multi-party communication complexity has also been widely investigated~\cite{dolev1989multiparty,sherstov2012multiparty}.
There are multiple reasonable models of multiparty communication studied in computer science; the number on the forehead~\cite{numberonforehead} model features players who get to see all inputs except their own while the number in the hand model only gives players access to their own inputs.
In addition to considerations about input, one can design multiparty communication models with different messaging systems.
The blackboard model~\cite{numberonforehead} has messages written on a shared blackboard that all players can read, while the message passing model~\cite{PVZ16} only lets players send messages directly to one another.
For instance, Ref.~\cite{viola2013} shows that for $k$ parties, each with an $n$-bit number (where $k$ is subpolynomial in $n$), determining if the sum of those numbers is some fixed value $s$ is possible using only $O(k\log k)$ bits of communication in the number in the hand model with direct message passing, albeit with a constant probability of error.

Another layer of complexity in distributed systems arises when privacy is essential: can agents jointly compute a function of their private inputs without revealing those inputs to one another? Foundational results demonstrate this is theoretically possible: Yao's garbled circuits protocol~\cite{yao1982protocols, yao1986generate} showed two parties can compute any function securely; the protocol of Ben-Or, Goldwasser, and Wigderson (BGW) extended this to multiple parties~\cite{ben2019completeness}. Closely related concepts include zero-knowledge proofs \cite{Kilian1992} and homomorphic encryption \cite{Acar2019}. These scenarios relate to various social processes, such as anonymous voting~\cite{chaum1981untraceable}, collaborative analysis of proprietary or sensitive data without disclosure~\cite{goldreich1998secure}, and public release of aggregated health statistics \cite{dyda2021differential}.

Many social systems can be seen as effectively performing distributed computing tasks. Further examples include (i) price discovery in markets~\cite{figuerola2010modelling}, which can be formalized as distributed consensus under strategic constraints, (ii) situations of multi-party bargaining~\cite{mcginn2012}---whether forming international alliances, corporate partnerships, or coalition governments (iii) establishment and maintenance of complex infrastructure such as transportation networks, power grids, and the internet, where no individual or institution is the sole arbiter~\cite{kovacs2003planning}. Theoretical results from the theory of distributed computing may provide insight into how these human systems operate, how coordination emerges or fails, and what mechanisms characterize resilient social institutions.

\subsection{Robustness and error correction}
\label{ssec:byz}

One major challenge arising in coordinated collective behavior broadly, and in distributed computing in particular, is the robustness of algorithms to local failures or adversarial behavior. A paradigmatic example is the Byzantine generals (BG) problem~\cite{lamport2019byzantine}, which captures the challenge of achieving reliable consensus when some participants may act maliciously or arbitrarily. Established results concern conditions for consensus given constraints on the number of faulty actors, communication network structure, and timing assumptions~\cite{lamport2019byzantine, vaidya2012iterative, castro1999practical, rabin1983}. Its analysis provides fundamental limits on collective computation: the impossibility result in~\cite{fischer1985impossibility} demonstrates that deterministic consensus is impossible in asynchronous systems even with a single faulty node; under a synchronous protocol, up to $\tfrac{n-1}{3}$ faulty nodes can be tolerated~\cite{lamport2019byzantine}. Many social phenomena involving collective behavior can be viewed as variants or generalizations of BG. For instance, game-theoretic bargaining with more than two agents \cite{bateni2010cooperative} and structured negotiation processes \cite{ortiz2002} map to extended BG problems. 

Another well-studied example in fault-tolerant computing is of distributed leader election and collective coin-flipping~\cite{russell1999lower}. Agents must agree on a common leader or random bit in spite of adversarial faults. These tasks have been widely studied under various assumptions~\cite{russell1999lower, balhara2014leader}. In addition to fault tolerance in distributed computing systems, robustness against errors in the accumulation of knowledge can be studied using other approaches. Recent work~\cite{brandenberger2025errors} investigates what strategy the authors of a scientific paper can use to correct for mistakes in previous papers on which their new paper depends. 

Numerous additional characterizations of robustness and error correction have arisen in computer science. For instance, self-stabilization~\cite{dijkstra1974self} demonstrates that systems can be designed to autonomously recover from perturbations. Many of these examples trace back to von~Neumann’s pioneering work on how a circuit can be reliable even when some of its components may be fault-prone~\cite{von1956probabilistic}. There is also a well-established theory about error-correcting codes \cite{macwilliams1977theory,pless1998introduction} which use redundancy to detect erroneous transmissions. These results formalize inherent trade-offs between synchrony assumptions, fault tolerance, and achievability of coordination. Extensions of classic BG scenarios have been considered, with varying network topology, synchrony assumptions, and incentives \cite{Litsas2014,correia2011,correia2010asynchronous,CORREIA20081291}; these cases more realistically emulate human organizational resilience.

\section{Related works}
\label{sec:related}

In this section, we briefly review earlier research related to the computational nature of human societies. We first describe several specific studies involving computational complexity in contexts related to human social systems (Sec.~\ref{ssec:related_complexity}). We then describe several past works that closely align with our Perspective (Sec.~\ref{ssec:related_papers}). 

\subsection{Computational complexity of social optima}
\label{ssec:related_complexity}

Game theory \cite{nisan2007introduction} and mechanism design \cite{jackson2014mechanism} investigate how to coordinate the actions of a collective, despite their often conflicting individual incentives, to achieve a ``social optimum'', often defined by equilibrium concepts~\cite{fudenberg1991game, arrow2024existence}. Individual preferences, strategies, and joint utilities are taken as given, and the resulting computational problem is to find the social optimum. Much of classical game theory and utility theory focus on the properties of equilibria~\cite{Osborne1994}, without evaluation of the computational tractability of finding or reaching them. Yet, the computational complexity of finding those social optima has been examined~\cite{papadimitriou2007complexity}, and the problem of finding Nash equilibria is believed to be computationally intractable~\cite{daskalakis2009complexity}. This difficulty is compounded when seeking a Nash equilibrium with certain properties, such as maximization of social welfare, often becoming $\mathsf{NP}$-hard~\cite{gilboa1989nash}. If there is no algorithm that finds the equilibrium in a reasonable amount of time, it is implausible that a group of individuals will converge to it.

On the other hand, if players coordinate their actions based on a mutually observed, external signal or trusted recommendation,there may be a {\it correlated equilibrium} \cite{JIANG2015347} that admits a polynomial time algorithm~\cite{papadimitriou2008computing}. In games prone to coordination failure, such as Prisoner's dilemma~\cite{flood1958some}, the correlated equilibrium often includes outcomes that are strictly Pareto superior (more efficient, in the game-theoretic sense \cite{duffy2010}) to all Nash equilibrium payoffs. The loss of this efficiency due to self-interested behavior is often stated as the {\it price of anarchy}~\cite{roughgarden2007introduction, johari2007price, christodoulou2005price, roughgarden2005selfish}. Therefore, in the presence of a coordination mechanism, the resulting equilibrium not only provides better joint utility but is also computationally easier to find. Efficiency is often taken to mean polynomial-time computability by a Turing~Machine, which takes the full payoff matrix as its input; however, this fails to capture the essential constraints imposed in systems with distributed information and computation.

Similar issues have been considered in algorithmic mechanism design, where the computational complexity of different mechanisms has been studied~\cite{nisan1999algorithmic}. A mechanism's role is to incentivize participants to truthfully reveal their private information; a central, trusted authority computes the incentives and decision. A mechanism is considered {\it strategy proof} if it is always in participants' best strategic interest to honestly portray their beliefs \cite{svensson1999strategy}. The information required to compute the function of interest is distributed across individuals; however, the computation itself is centralized. This framework applies to scenarios where trusted, centralized intermediaries naturally exist, such as government-run auctions and elections~\cite{papadimitriou2008hardness}, but has limited relevance in scenarios where a trusted central authority is absent. Examples include free markets and peer-to-peer networks~\cite{feigenbaum2004distributed}. In addition, the role of strategic behavior, when individuals performing the computation in a distributed setting prefer certain outcomes over others, remains largely unexplored.

In~\cite{nisan2006communication}, the authors study the communication complexity of determining prices of certain kinds of indivisible goods with combinatorial preferences. They show that achieving optimal allocation using prices requires exponential (in the number of goods) communication, and prices must be discovered for all possible combinations of items. In practice, this computational intractability makes price discovery infeasible for such goods. However, for goods that don't require combinatorial preferences, the price of a bundle simply equals the sum of individual item prices, and using prices for allocation is computationally efficient. So, whether prices can efficiently coordinate allocation depends fundamentally on the structure of preferences.

Computational social choice theory constitutes a complementary line of inquiry~\cite{brandt2016handbook, chevaleyre2007short, bartholdi1989computational, bartholdi1989voting} that studies the algorithmic and computational complexity of aggregating individual preferences into collective outcomes. Social choice theory concerns identifying aggregation mechanisms that satisfy certain desirable properties~\cite{arrow1964social}. A parallel line of work highlights the strategic manipulation of such mechanisms, showing that many desirable rules are inherently susceptible to manipulation, in the sense that individuals or coalitions can alter outcomes by misreporting their preferences~\cite{gibbard1973manipulation, satterthwaite1975strategy}. Computational social choice extends these investigations by incorporating tools from theoretical computer science to ask whether such strategic behavior is computationally feasible in practice, for example, by analyzing the complexity of manipulation, control, or bribery under different voting rules.

\subsection{Closely aligned existing works}
\label{ssec:related_papers}

We now survey a couple of lines of research that genuinely embody the type of computational viewpoint we espouse herein.

A recent exemplary work that our Perspective aligns with is Ref.~\cite{hebert2025governance}, in which {\it governance} is cast as a complex satisfiability problem. They instantiate this view of governance in a concrete social hypergraph model, allowing direct simulation in synthetic social systems. Examining a range of governance structures from centralized to distributed (as discussed in Sec.~\ref{ssec:coordination}), they find efficient solution-finding by intermediately decentralized governance; specifically, wherein, as they write, ``{\it small overlapping decision groups make specific decisions and share information},'' closely aligning with notions of distributed computing (Sec.~\ref{ssec:distributed}). They introduce the notion of a ``democratic satisfiability problem'' for which solutions should not only satisfy the given constraints but also closely reflect the distribution of preferences in the population. We note that these authors consider the {\it fraction} of constraints that are satisfied rather than focusing on achievement of perfect solutions, akin to using approximation complexity over worst-case complexity (Sec.~\ref{sssec:approx}).

Notable work has examined the role of computational complexity in individual cognition. For instance, everyday cognitive tasks often map onto $\mathsf{NP}$-hard problems---e.g., planning, inference, causal reasoning \cite{ROTH1996273,koller2009probabilistic}---yet arguably, people solve them quickly and frequently. How do bounded agents manage when faced with computationally intractability? To address this, van Rooij \cite{vanRooij2019cognition} introduces parameterized complexity to cognitive science. Parameterized complexity partitions problem classes into tractable subsets \cite{Niedermeier2006,downey2013fundamentals}. They argue that human cognition doesn't solve the full problem but identifies and operates within tractable partitions, e.g., via domain-specific simplifications. This work closely aligns with our Perspective, in that not only does it apply reasoning from computer science theory to human systems, it also heavily relies on concepts beyond the traditional TM and worst-case complexity---for instance, fixed-parameter tractability, which describes problems that are efficiently solvable given bounded parameters, even if the general problem is intractable.

\subsection{Clarifications of scope}
\label{ssec:related_scope}
It is important to clarify how the research agenda we present in this Perspective differs from several distinct yet thematically similar lines of research.

We first differentiate from the established field of {\it computational social science} \cite{lazer2009css,lazer2020css}, commonly understood as the application of computational and data-driven methods to empirical social science questions \cite{conte2012css}. This field is enabled by the increasing digitization of social life, and has substantially expanded the scope and focus of the social sciences, particularly through the analysis of digital trace data from online platforms and communication networks \cite{salganik2017bitbybit,lazer2020css}. Computational social science also overlaps with computational {\it modeling} of social systems, via, e.g., agent-based models \cite{squazzoni2012agent}, cellular automata \cite{wolfram1984cellular}, random graphs \cite{van2024random}, iterated games \cite{tadelis2013game}, and related methods---approaches that are more similar to our objective. However, the aim of this Perspective is conceptually distinct from those works. Rather than applying computational tools and models to social data, we focus on how social systems themselves can be understood as computational.

We also distinguish from research into the widespread deployment of algorithms from computer science in society, underpinning much of modern digitized technology~\cite{Tardieu2020}. Algorithms for searching, sorting, encryption, and countless other functions, have led to rapid transitions in societal information processing \cite{beer2017social}; for the modern era, a computational viewpoint becomes unavoidable. Numerous critical issues arise such as algorithmic fairness, transparency, and data ethics \cite{wang2022brief,yang2023algorithmic,Olhede}. An understanding of these issues and the increasing ubiquity of algorithms and how humans interact with them can certainly provide important insights into modern social systems \cite{Seaver2017}. However, our Perspective is advocating for a descriptive view of social processes themselves as computational, as opposed to characterizing how social systems literally employ digital algorithms~\cite{kaczmarczyk2016computers}. Beyond the modern era, we are interested in how the computational lens can provide insight into human social systems throughout history.

Finally, there is much research into development and analysis of algorithms that are motivated by practical utility, such as novel voting mechanisms \cite{Alturki2016,neff2001verifiable,yang2018multiwinner,Sako1994}, encryption schemes \cite{dixit2017traditional}, autonomous vehicles and robotics \cite{klavins2004communication,ming2021survey}, and artificial intelligence systems \cite{Luger2025}, and more broadly, research into the design and implementation of algorithms for societal benefit \cite{kaczmarczyk2016computers}. More boundary-pushing directions include schemes for wholesale integration of computational and social systems \cite{evans2020social}, fully algorithmic decentralized governance structures \cite{Just2017}, and AI alignment tasks to mitigate risks of superintelligence \cite{Ji2025,soares2014aligning}. These questions are computational in character, and have important societal implications for the near and far future. However, in this Perspective, we emphasize how computer science theory could help to understand social systems, rather than how it could help design or intervene on them.

\section{Conclusion and future directions}
\label{sec:conclusion}

Human societies routinely confront large-scale coordination problems and information processing tasks. Some of these problems are relatively easy to manage, while others are profoundly difficult. With the difficulty depending not only on the scale of the society but also on the organizational structures and mechanisms through which coordination is achieved. What has not traditionally been examined is the {\it computational} nature of these social processes: the resources they require, how hard they are to solve or even approximate, and how these constraints, in turn, shape the organization and evolution of social institutions; this Perspective is motivated by the prospect of addressing this gap.

Understanding these processes requires integrating and extending computational complexity theory with the empirical study of social organization. While computational complexity theory provides rigorous tools for analyzing resource constraints, it remains a forefront research challenge to formally capture many of the features arising in human groups (Sec.~\ref{sec:attributes}). These features differ in many ways from the centralized models of computation such as the Turing~Machine (Sec.~\ref{ssec:TM}); this has motivated our exposition of the distributed computing paradigm (Sec.~\ref{ssec:distributed}), non-uniform computation (Sec.~\ref{ssec:circuits}), and notions such as approximation- and average-case complexity (Sec.~\ref{ssec:complexity_notions}). We anticipate that further analysis of the computational aspects of human social systems will raise many interesting questions and challenges for theoretical computer science.

We have suggested a variety of directions for fruitful future research, primarily in two categories---the development of new computer science theory as inspired by human collective computation, and the analysis of human social phenomena through the lens of computation. Some examples of prospects in the former category include complexity characterizations of price discovery in {\it distributed} systems, extending previous works in settings of centralized computation~\cite{Scarf1982,Papadimitriou2010,Codenotti2005,roughgarden2010computing}; another example is formulation of computational complexity notions that ask, rather than how much computation is required to solve or approximate the solution of a problem, how well can a solution be approximated under fixed constraints of finite time, memory, or energy budget. In the latter category, we suggest examination of real-world human social coordination tasks through the lens of computer science, allowing determination of {\it how difficult} these collective problem solving tasks are.

More broadly, many longstanding debates in economics, sociology, political science, anthropology, and related fields could be placed on concrete and formal footing within a computational framework. From hunter-gatherer societies to modern states and markets, such a framework could help explain conditions for efficiency of free markets versus centralized mechanisms (Sec.~\ref{ssec:coordination}), why hierarchy and modularity often emerge (Sec.~\ref{ssec:mod_hier}), why social systems restructure themselves with scale (Sec.~\ref{ssec:scaling}), and how external memory systems provide benefits for information processing (Sec.~\ref{ssec:memory}), among others. We have spelled out some elements of what such a framework could look like, but we have not directly proposed one herein, avoiding hastened over-formalization. Even in the absence of a rigorous formalism, we hope this work inspires broader adoption of the computational lens as a qualitative, explanatory viewpoint in the study of social systems.

\ \\

\section*{Acknowledgments}
MJH, DHW, AY, HH, and AJS thank the Santa Fe Institute for support.
JK acknowledges support from the Austrian Science Fund (FWF) under Grants No. 10.55776/P34994 and EFP5 ReMASS, funding from the Austrian Federal Ministry for Climate Action, Environment, Energy, Mobility, Innovation, and Technology under GZ 2023-0.841.266, through the Postdoc Program for Complexity Science and Data Competence.

NK's research was supported by a Schmidt Sciences Polymath award to David Soloveichik and the Laboratory Directed Research and Development Program at Sandia National Laboratories. His contributions to this work started when he was attending The University of Texas at Austin.

This article has been authored in part by an employee of National Technology \& Engineering Solutions of Sandia, LLC under Contract No. DE-NA0003525 with the U.S. Department of Energy (DOE). The employee owns all right, title and interest in and to the article and is solely responsible for its contents. The United States Government retains and the publisher, by accepting the article for publication, acknowledges that the United States Government retains a non-exclusive, paid-up, irrevocable, world-wide license to publish or reproduce the published form of this article or allow others to do so, for United States Government purposes. The DOE will provide public access to these results of federally sponsored research in accordance with the DOE Public Access Plan \href{https://www.energy.gov/downloads/doe-public-access-plan}{https://www.energy.gov/downloads/doe-public-access-plan}.

\bibliography{comp.bib}

\end{document}